\documentstyle[aps]{revtex}

\begin{document}

\title{How many family-independent, independent U(1)'s, can we gauge?}
\author{William A. Ponce\thanks{e-mail wponce@fisica.udea.edu.co}, 
John F. Zapata and Daniel E. Jaramillo\\
Depto. de F\'\i sica, Universidad de Antioquia,\\
A.A. 1226, Medell\'\i n, Colombia.}

\maketitle

\abstract{In the context of $SU(3)_c\otimes SU(2)_L\otimes [U(1)]^n$, 
$SU(3)_c\otimes SU(2)_L\otimes SU(2)_R\otimes [U(1)]^{n'}$ and 
$SU(3)_c\otimes SU(3)_L\otimes [U(1)]^{n''}$, we analyze the possible values 
for $n, n'$ and $n''$. That is, we look for the number of family-independent,
independent  abelian hypercharges that can be gauged in each one of those
models. We find  that $n=n'=1$ for the minimal fermion content, but $n$ can
take any value  if we add the right-handed neutrino field to each family in the
Standard  Model.}

\section{Introduction}
The standard model (SM) local gauge group 
$SU(3)_c\otimes SU(2)_L\otimes U(1)_Y\equiv G_{SM}$ with 
$SU(2)_L\otimes U(1)_Y$ hidden
and $SU(3)_c$ confined, has been very sucessfull in explaining a huge amount
of experimental data related with particle interactions at low energies.
Nevertheless, phycisists have been looking for variants of this model in order 
to explain several experimental facts such as parity violation, the 
mass spectrum of the known elementary fermions, 
or the explanation of different anomalies, in the context of extended 
electroweak theories as for example $SU(2)_L\otimes SU(2)_R\otimes 
U(1)_{(B-L)}$ or $SU(2)_L\otimes U(1)_X\otimes U(1)_Z$, etc..

More profound questions such as why charge is quantized\cite{babu} or why 
the low energy gauge group is the one belonging to the SM\cite{ponce} 
are still unanswered. In this note we are going to refer to a 
different fundamental question which is related to the number of 
family-independent, independent 
local gauge Abelian groups that can be gauged simultaneously.

\section{Standard Model}
First let us start our analysis in the context of a family-independent 
$U(1)$ Abelian factor and look for the constraints that anomaly cancellation 
imposes on such a model in an environment provided by the SM gauge group 
$SU(3)_c\otimes SU(2)_L\otimes U(1)_Y$. The anomaly cancellation constraint 
equations are\cite{babu}
\begin{eqnarray}\label{ano1}
[SU(2)_L]^2U(1) &:& 3Y_q+Y_f=0\\ \label{ano2}
[SU(3)_c]^2U(1) &:& 2Y_q+Y_u+Y_d=0\\ \label{ano3}
[grav]^2U(1)  &:& 6Y_q+3Y_u+3Y_d+2Y_f+Y_e=0\\ \label{ano4}
[U(1)]^3      &:& 6Y_q^3+3Y_u^3+3Y_d^3+2Y_f^3+Y_e^3=0
\end{eqnarray}
where $Y_\eta(\eta=q,f,u,d,e)$ are the $U(1)_Y$ hypercharges of the 
following multiplets: $\psi_q^T=(u,d)_L\sim [3,2],\;\psi_f^T=(\nu,e^-)_L
\sim [1,2],\;\psi_u=u_L^c\sim [\bar{3},1],\;\psi_d=d_L^c\sim [\bar{3},1]$ 
and $\psi_e=e_L^+\sim [1,1]$ which are the fifteen states belonging to one 
family (the upper $c$ symbol stands for charge conjugation and the 
numbers in the square bracket stand for the $[SU(3)_c,SU(2)_L]$ quantum 
numbers).

By combining the cubic equation with the other three lineal ones we get:
\begin{equation}\label{cub}
Y_q(2Y_q-Y_u)(4Y_q+Y_u)=0,
\end{equation}
\noindent
which provides three independent families of solutions 
as quoted in Table I, where the SM hypercharge value has been included for 
comparison.

\vspace{.5cm}

\begin{center}
\begin{tabular}{||l||c|c|c|c|c||} \hline\hline
Sol.    & $Y_q$ & $Y_f$ & $Y_e$ & $Y_u$ & $Y_d$ \\ \hline\hline
{\bf A} & 0     & 0     & 0     & a     & $-$a \\
{\bf B} & b     & $-3$b & 6b    & 2b    & $-4$b \\
{\bf C} & c     & $-3$c & 6c    & $-4$c & 2c \\ \hline
{\bf SM}& 1/3   & $-1$  &  2    & $-4/3$& 2/3 \\ \hline\hline
\end{tabular}
\end{center}

\vspace{.5cm}

\noindent
In Table I $a,b$ and $c$ are real arbitrary parameters, and the SM 
hypercharge is related with solution {\bf C}, for $c=1/3$. 

From Table I we realize that charge is 
not quantized in the context of the SM\cite{babu}. Not only there are 
three different solutions to the anomaly constraint equations, but 
we do not know why solution {\bf C} is the one chosen by nature, and 
even worse, we can not explain why $c=1/3$ for that particular 
solution\cite{babu}.

Now, taking into account all three solutions, namely $U(1)_{Y_A}$, 
$U(1)_{Y_B}$ and $U(1)_{Y_C}$  one can ask whether 
$U(1)_{Y_A}\otimes U(1)_{Y_B}\otimes U(1)_{Y_C}$ can be gauged 
simultaneous, or at least two of them at the same time. In order to answer 
the second question we must look for solution to the following two new 
anomaly constraint equations:
\begin{equation}\label{e21}
6Y_{q\alpha}^2Y_{q\beta}+3Y_{u\alpha}^2Y_{u\beta}+3Y_{d\alpha}^2Y_{d\beta}+
2Y_{f\alpha}^2Y_{f\beta}+Y_{e\alpha}^2Y_{e\beta}=0
\end{equation}

\begin{equation}\label{e12}
6Y_{q\alpha}Y_{q\beta}^2+3Y_{u\alpha}Y_{u\beta}^2+3Y_{d\alpha}Y_{d\beta}^2+
2Y_{f\alpha}Y_{f\beta}^2+Y_{e\alpha}Y_{e\beta}^2=0
\end{equation}
for $\alpha ,\beta={\bf A,B,C}$. The only solution to Eqs.(\ref{e21}) and 
(\ref{e12}) is for $\alpha=\beta$, with $a,b$ and $c$ still free parameters. 
So, only one solution to the anomaly 
constraint equations can be gauged, and once it has been gaugen it can be 
gauged as many times as we want; but we can not gauge two different 
solutions at the same time\cite{pon}.

From the previous analysis we conclude that the hypercharges $Y_A$ and 
$Y_B$ should be automatically excluded once $Y_C$, with $c=1/3$, is gauged 
as the correct hypercharge chosen by nature. So, in the context of 
$SU(3)_c\otimes SU(2)_L\otimes U(1)_Y$ there is only one anomaly free 
independent hypercharge corresponding to $Y=Y_C$ with $c=1/3$. That 
hypercharge corresponds to the SM hypercharge, and it can be gauged as many 
times as we wishes allowing the value for $c$ to be a free parameter.

\section{SM plus right-handed neutrino}
When the right-handed neutrino $\psi_\nu=\nu^c_L\sim [1,1]$ is included 
in each family as part of the spectrum, the set of anomaly 
constraint equations become:
\begin{eqnarray}\label{nano1}
[SU(2)_L]^2U(1) &:& 3Y_q+Y_f=0\\ \label{nano2}
[SU(3)_c]^2U(1) &:& 2Y_q+Y_u+Y_d=0\\ \label{nano3}
[grav]^2U(1)  &:& 6Y_q+3Y_u+3Y_d+2Y_f+Y_e+Y_\nu=0\\ \label{nano4}
[U(1)]^3      &:& 6Y_q^3+3Y_u^3+3Y_d^3+2Y_f^3+Y_e^3+Y_\nu^3=0
\end{eqnarray}
where $Y_\nu$ is the $U(1)_Y$ hypercharge associated with the right-handed 
neutrino field.
Combining again the cubic equation with the three lineal ones we get:
\begin{equation}\label{cubn}
Y_q(2Y_q-Y_u-Y_\nu)(4Y_q+Y_u-Y_\nu)=0,
\end{equation}
\noindent
which again provide a set of three independent families of solutions as 
presented in Table II (notice by the way that Eq. (\ref{cubn}) reduces to 
Eq. (\ref{cub}) for $Y_\nu =0$ as it should).

\vspace{.5cm}

\begin{center}
\begin{tabular}{||l||c|c|c|c|c|c||} \hline\hline
Sol. & $Y_q$ & $Y_f$ & $Y_e$ & $Y_\nu$ & $Y_u$ & $Y_d$ \\ \hline\hline
{\bf A'} & 0 & 0 & $\eta$ & $-\eta$ & $\delta$ & $-\delta$ \\
{\bf B'} & (a+b)/2 & $-3$(a+b)/2 & 3a+2b & b & a & $-$(2a+b) \\
{\bf C'} & $-$(A$-$B)/4 & 3(A$-$B)/4 & (B$-$3A)/2 & B & A & $-$(A+B)/2 \\
\hline\hline
\end{tabular}
\end{center}

\vspace{.5cm}

\noindent
where $\eta ,\delta , a,b,A$ and $B$ are real arbitrary parameters. Notice 
that solution {\bf B'} is independent of solution {\bf A'} for $b\neq -a$ 
and solution {\bf C'} is independent of solution {\bf A'} for $B\neq A$.
From Table II we can see that the SM hypercharge value is related with 
solution {\bf C'} for $B=0$ and $A=-4/3$, and the $B-L$ 
(Baryon number minus Lepton number) 
hypercharge is related with solution {\bf B'} for $a=-1/3$ and 
$b=1$. Again, charge is not quantized in this simple extension of the SM.

But can we now gauge $U(1)_{Y_{A'}}\otimes U(1)_{Y_{B'}}\otimes 
U(1)_{Y_{C'}}$ simultaneously? Or at least two of them at the same time? To 
answer the second question we must look for solutions to the following two 
new anomaly constraint equations

\begin{equation}\label{en21}
6Y_{q\alpha}^2Y_{q\beta}+3Y_{u\alpha}^2Y_{u\beta}+3Y_{d\alpha}^2Y_{d\beta}+
2Y_{f\alpha}^2Y_{f\beta}+Y_{e\alpha}^2Y_{e\beta}+ 
Y_{\nu\alpha}^2Y_{\nu\beta}=0
\end{equation}

\begin{equation}\label{en12}
6Y_{q\alpha}Y_{q\beta}^2+3Y_{u\alpha}Y_{u\beta}^2+3Y_{d\alpha}Y_{d\beta}^2+
2Y_{f\alpha}Y_{f\beta}^2+Y_{e\alpha}Y_{e\beta}^2
+Y_{\nu\alpha}Y_{\nu\beta}^2=0,
\end{equation}
for $\alpha ,\beta=$ {\bf A',B'} and {\bf C'}. Solutions to this set of 
equations for the three possible combinations of values for $\alpha$ and 
$\beta$ are presented in Table III.

\hspace{.5cm}

\begin{center}
\begin{tabular}{||c||c|c|c||} \hline\hline
   &    {\bf A'}  &  {\bf B'} & {\bf C'} \\ \hline\hline
{\bf A'} & Trivial & $b=-a; \;\; \eta ,\delta$ arbitrary (Trivial) & 
$B=A; \;\; \eta ,\delta$ arbitrary (Trivial) \\
  &  &  $\eta=\delta ,\;\; a,b$ arbitrary & $\eta=-\delta ,\;\; A,B$ arbitrary \\  \hline
{\bf B'} & & Trivial & $B=-3A, \;\; a,b$ arbitrary \\
  &  &  & $b=-3a, \;\; A,B$ arbitrary \\ \hline
{\bf C'}  &  &   &    Trivial \\ \hline\hline
\end{tabular}
\end{center}

\hspace{.5cm}

\noindent
From Table III we see that we can gauge one of the three solutions at 
least two times (what we call the 
Trivial solutions in the Table), but what is more important now is that, 
contrary to the SM situation, we can gauge any two different solutions at 
the same time.

But, can we gauge the three solutions at the same time? To answer this 
question the further anomaly constraint equation must also be satisfied
\begin{eqnarray}\label{en12e}
6Y_{qA'}Y_{qB'}Y_{qC'}+3Y_{uA'}Y_{uB'}Y_{uC'}+3Y_{dA'}Y_{dB'}Y_{C'}\hspace{1cm}\nonumber\\
+2Y_{fA'}Y_{fB'}Y_{C'}+Y_{eA'}Y_{eB'}Y_{C'}+Y_{\nu A'}Y_{\nu B'}Y_{\nu C'}=0.
\end{eqnarray}
The algebra shows now the following solutions: $A'=B'=C'$ (Trivial), or 
$A'=B'$, or $A'=C'$ or $B'=C'$. So, we can gauge one of the three familes of 
solutions three or more times (in fact as many times as we wishes), which in
turn  provides an infinity number of independent $U(1)'s$ that can be gauged 
simultaneously; also we can gauge two independent families at the same time, 
but we can not gauge the three independent families simultaneously.

Our conclusion is now that once we gauge solution {\bf C'} for $B=0$ and 
$A=-4/3$ (the SM hypercharge value) we can gauge again solution {\bf C'} 
as many times as we want with $A$ and $B$ free parameters, and still we can 
gauge either solutions {\bf A'} for $\eta=-\delta$ or either solution 
{\bf B'} for $b=-3a$. In particular we can gauge the hypercharge $B-L$
simultaneously with the SM hypercharge $Y$.  This new amazing result is a
consequence of introducing the right-handed  neutrino field and it is a
novelty not present in the SM.

\section{The Left-Right symmetric model}
The Left-Right symmetric model of the electroweak interactions was introduced 
in the middle of the seventies\cite{pati} in order to elucidate the origin of 
parity violation in low energy physics. Within the framework of local gauge 
theories the idea was to enlarge the SM local gauge group $G_{SM}$ to the 
Left-Right symmetric one $SU(3)_c\otimes SU(2)_L\otimes SU(2)_R\otimes 
U(1)_X\equiv G_{LR}$, where in the literature $X=B-L$\cite{pati}.

The right-handed neutrino is automatically included as an isopartner of $e_R$ 
in $G_{LR}$. The anomaly constraint equations on $X$ are now:

\begin{eqnarray}\nonumber
[SU(2)_L]^2U(1) &:& 3X_{qL}+X_{fL}=0\\ \nonumber
[SU(2)_R]^2U(1) &:& 3X_{qR}+X_{fR}=0\\ \nonumber
[SU(3)_c]^2U(1) &:& X_{qL}-X_{qR}=0 \\ \nonumber
[grav]^2U(1)  &:& 6X_{qL}-6X_{qR}+2X_{fL}-2X_{fR}=0\\ \nonumber
[U(1)]^3  &:& 6X_{qL}^3-6X_{qR}^3+2X_{fL}^3-2X_{fR}^3=0
\end{eqnarray}
where $X_\eta (\eta=qL,qR,fL,fR)$ are the $U(1)_X$ hypercharges of the 
following multiplets: $\psi^T_{qL}=(u,d)_L,\; \psi^T_{qR}=(u,d)_R,\; 
\psi^T_{fL}=(\nu ,e)_L,\; \psi^T_{fR}=(\nu ,e)_R$.

The above equations imply the unique solution $X_{qR}=X_{qL}$ and 
$X_{fL}=X_{fR}=-3X_{qL}$ (i.e. the $U(1)_X$ is vectorlike). So, there is 
only one independent family of hypercharge solutions in $G_{LR}$ (for 
$X_{qL}=1/3$ it is the $B-L$ hypercharge). Again, there is not charge 
quantization in $G_{LR}$ because $X_{qL}$ is a free parameter.

Now let us analyze the group $G_{LR}\otimes U(1)_Z$. The new anomaly 
constraint equations to be considered are: 
\begin{eqnarray} \nonumber
[U(1)_X]^2U(1)_Z  &:& 6X_{qL}^2Z_{qL}-6X_{qR}^2Z_{qR}+2X_{fL}^2Z_{fL}
-2X_{fR}^2Z_{fR}=0 \\ \nonumber
[U(1)_Z]^2U(1)_X  &:& 6Z_{qL}^2X_{qL}-6Z_{qR}^2X_{qR}+2Z_{fL}^2X_{fL}
-2Z_{fR}^2X_{fR}=0
\end{eqnarray}
The solution to the former set of equations (with $Z_{qR}=Z_{qL}$ and 
$Z_{fL}=Z_{fR}=-3Z_{qL}$) is only the trivial one $X_\eta=Z_\eta$ for 
$\eta=qL,qR,fL,fR$. Again, we can gauge two or more $U(1)'s$  in the
context of $SU(2)_L\otimes SU(2)_R\otimes SU(3)_c$, as far the  hypercharge of
those $U(1)'s$ are all of them proportional to $B-L$ and,  as in the SM case,
they are not independent.

\section{The (3,3,1) model}
Finally let us consider the gauge group $SU(3)_L\otimes SU(3)_c\otimes 
U(1)_K\equiv G_{331}$ as an extension of the SM group\cite{pleitez}, and let us
study  the anomaly constraint equations for one single family of quarks and 
leptons as it is done for the SM.

For one family let us start with $\psi_L^T=(u,d,q)_L$ with  
$G_{331}$ quantum numbers $(3,3,K_Q)$, where $q_L$ is an exotic quark 
of electric charge to be fixed ahead. Now, in order to have $SU(3)_c$ 
vectorlike as in the SM we must introduce the fields 
$\psi_u=u_L^c,\; \psi_d=d_L^c$ and $\psi_q=q_L^c$ with $G_{331}$ quantum 
numbers given by $(1,\bar{3},K_u),(1,\bar{3},K_d)$ and $(1,\bar{3},K_q)$ 
respectively.
Next, in order to introduce the lepton fields and to cancel the $SU(3)_L$ 
anomaly we must define three new multiplets $\chi_{1L}^T=(e^-,\nu ,l_1)_L, 
\; \chi_{2L}^T=(l_2,l_3,l_4)_L$ and $\chi_{3L}^T=(l_5,l_6,l_7)_L$ with 
$G_{331}$ quantum numbers $(\bar{3},1,K_1),\;(\bar{3},1,K_2)$ and 
$(\bar{3},1,K_3)$ respectively, where $l_i,\; i=1,...7$ are related to 
lepton fields, most of them exotics.

The anomaly constraint equations on $K$ are now:
\begin{eqnarray} \nonumber
[SU(3)_c]^2U(1) &:& 3K_Q+K_u+K_d+K_q=0\\ \nonumber
[SU(3)_L]^2U(1) &:& 3K_Q+K_1+K_2+K_3=0\\ \nonumber
[grav]^2U(1)  &:& 9K_Q+3K_u+3K_d+3K_q+3K_1+3K_2+3K_3=0\\ \nonumber
[U(1)]^3      &:& 9K_Q^3+3K_u^3+3K_d^3+3K_q^3+3K_1^3+3K_2^3+3K_3^3=0.
\end{eqnarray}
The former set of equations imply $K_Q=0$ and $K_uK_dK_q=-K_1K_2K_3$ which 
in turn imply an infinity set of independent solutions. 
In order to reduce the number of 
independent solutions and to construct a reasonable model, further constraints 
must be introduced, so let us look for solutions to the former set of 
equations such that $K_q=K_d$ and $K_1=K_2$ (notice that the anomaly
constraint equations on $K$ are $K_u\leftrightarrow K_d$ and
$K_1\leftrightarrow K_2$ symmetric); with these conditions we  warrant that not
exotic electric charges are present on the model; using them  we get 
$K_q=K_d\equiv K=-K_u/2=-K_1=-K_2=K_3/2$. Again, $K$ is a free  parameter and
charge is not quantized.

Using for the symmetry breaking chain $SU(3)_L\longrightarrow SU(2)_L\otimes 
U(1)_S$ the branching rule $3\longrightarrow 2(s)+1(-2s)$, and the electric 
charge generator defined by $Q=T_3+S+K$ where $T_3=\pm 1/2$ is the 
third isospin component, we have for $s=1/6$ the following electric 
charges: $Q_u=2/3, \; Q_d=Q_q=-1/3$ and $\chi_{1L}^T=(e^-,\nu ,N_1^0)_L, 
\;\chi_{2L}^T=(E^-,N_2^0,N_3^0)_L$ and $\chi_{3L}^T=(N_4^0,E^+,e^+)_L$, 
where $N_i^0,\; i=1,2,3,4$ are four exotic neutrino fields and $E$ is an
exotic  electron. The 27 fields in $\psi_L,\psi_u,\psi_d,\psi_q$ and
$\chi_{iL},\;  i=1,2,3$ are nothing else but the 27 fields in the fundamental 
representation of the grand unification group $E_6$\cite{e6}. As a matter 
of fact, the group representation that we have derived in this exercise is
such that  $G_{331}\subset E_6$\cite{sanchez}, but of course there are many
possible realistic models in the context of the gauge group
$G_{331}$\cite{pleitez}.

\section{Flipped $SU(5)\otimes U(1)$}
In this section we analyzed the flipped $SU(5)\otimes U(1)_W$ model\cite{su5} 
as an application of the formalism presented in Section 3. This model 
unifies the 16 fields in one family at a high scale, and yet has an extra 
$U(1)$ quantum number which does not coincide with the SM hypercharge $Y$. 
In this model the minimal choice of multiplets that are free of the 
$[SU(5)]^3$ anomaly are\cite{su5} $(10,\; W_{10}),\; (\bar{5},\; W_5)$ and 
$(1,\; W_1)$. The other anomaly constraint equations are:
\begin{eqnarray}
[SU(5)]^2U(1) &:& W_5+3W_{10}=0\\ \nonumber
[grav]^2U(1)  &:& 5W_5+10W_{10}+W_1=0\\ \nonumber
[U(1)]^3      &:& 5W_5^3+10W_{10}^3+W_1^3=0,
\end{eqnarray}
which imply $W_{10}=-W_5/3=W_1/5$ with $W_{10}$ a free parameter (again 
charge is not quantized in this model either). 

A simple check shows that this $U(1)_W$ is just a particular case of 
solution {\bf C'} in the Table II in Section 3, with $A=W_{10}$ and 
$B=5W_{10}$. (This same $U(1)_W$ is the one present in the grand unification 
group $SO(10)$\cite{so10} for the breaking chain 
$SO(10)\longrightarrow SU(5)\otimes U(1)_W$ which has the branching rule 
$16\longrightarrow 1(-5)+\bar{5}(3)+10(-1)$.)

\section{ACKNOWLEDGMENTS}
This work was partially supported by Colciencias and BID in Colombia.

\end{document}